# Compressive fluorescence spectral imaging with a spectrometer *

Chao Wang(王超)[1,2,3], Xue-Feng Liu(刘雪峰)[3†], Wen-Kai Yu(俞文凯)[1], Xu-Ri Yao(姚旭日)[3], Fu Zheng(郑福)[3], Qian Dong(董乾)[3], Ruo-Ming Lan(蓝若明)[4], Guang-Jie Zhai(翟光杰)[3], and Qing Zhao(赵清)[1]

[1] *School of Physics, Beijing Institute of Technology, Beijing 100081,China*
[2] *China Academy of Engineering Physics, Mianyang 621900, China*
[3] *Key Laboratory of Electronics and Information Technology for Space Systems, National Space Science Center, Chinese Academy of Sciences, Beijing 100190, China*
[4] *School of Physics and Electronics, Shandong Normal University, Jinan 250014, China*

**Abstract**

We present an efficient approach and principle experiment for compressive sensing (CS) fluorescence spectral imaging. According to the dimension-reduced effect of CS, the spectral and spatial information was simultaneously obtained by using a fiber spectrometer without mechanical scanning. As a method verification, we demonstrated spectral imaging under only two typical wavelengths, but the spectral resolution is up to 1.4nm depended on the fiber spectrometer. The method could obtain 50% light energy from the object, much larger compared with mechanical scanning which detects light of only one point per measurement. The relationship between sampling rate and image quality is also discussed in this study.

## 1. Introduction

Fluorescence imaging is a common method in basic and applied scientific researches, and is employed in analyzing the special chemical composition and physical structure with microscopy.[1] Spectral imaging can be generally defined as the combined acquisition of spatial and spectral information. Imaging spectrometers are also mounted airborne and in satellites for remote sensing and astronomical observation.[2-4] Spectral information can be used in matter analysis in many fields.[5-7] In biomedical research, an extensive range of applications such as research of protein localization and interactions requires quantitative approaches to analyze several distinct fluorescent molecules at simultaneous time in the same sample.[8] In fact, these applications are becoming increasingly common with the availability of various fluorescent dyes and proteins with emission ranging from UV to far infrared.[9] Fluorescence spectral imaging technology has, indeed, become a fundamental tool for scientific research.[10-14]

---

* Project supported by the National Major Scientific Instruments Development Project of China [Grant No. 2013YQ030595], and the National Natural Science Foundation of China [Grant No. 61601442, 11275024 and 11675014], Science and Technology Innovation Foundation of Chinese Academy of Sciences [CXJJ-16S047].
† Corresponding author. E-mail: liuxuefeng@nssc.ac.cn

However, some disadvantages exist in the traditional spectral imaging. In most current spectral imaging technologies, the spatial information is acquired by mechanically scanning the sample point by point using a spectrometer. Inevitably, mechanical movement will produce errors in spatial domain, potentially requiring measurement repeats which waste resources. Additionally, spectral imaging has a remarkable feature of large amount of data, so it is often highly compressible.

In digital signal processing area, compressive sensing (CS) has become a popular field since Donoho published the "Compressive sensing" report based on the mathematical theory in the last century.[11] Since then, in 2008, Candes and Wakin provided an introduction to compressive sampling[12] and Baraniuk et al. developed a new imaging approach using a single-pixel camera based on the theory of CS. This is the first implementation of CS for optical imaging.[15] Compared to Shannon-Nyquist sampling theorem, CS provides a sensing framework sampling sparse signals in a more efficient way. The compressed signal is obtained during acquisition with CS, thus avoiding the requirement of digital compression. Since CS requires fewer measurements, it is highly suitable for spectral imaging. Single-pixel camera has also been used in spectral imaging which formerly needs a 2D array detector to obtain the spectral image of an object.[16-24]

In this work, we focus on fluorescence spectral imaging using single-pixel camera in which the detector is a spectrometer instead of the traditional 2D array detector. This imaging modality can benefit from CS, since the fluorescence spectral data is typically compressible. The number of measurements is reduced and the imaging process does not require mechanical scanning.

## 2. Compressive sensing

As defined by Shannon-Nyquist sampling theorem, a limited bandwidth signal can be reconstructed by acquiring the original signal at a rate of at least twice the bandwidth. This provides a theoretical basis for many applications. However, most of nature signals have a sparse representation in a certain basis. Therefore, we can reconstruct this type of signal with CS theory beyond Shannon-Nyquist sampling theorem. Supposing $x$ is the target signal of $n$ dimensions, instead of measuring the target signal itself, we can measure the signal modulated with matrix $A$:

$$y = Ax + e \tag{1}$$

where $y$ is the measurement value with m-dimension and $e$ is the measurement noise. The matrix $A$ has dimensions of $m \times n$ with $m < n$. Understandably, the equations are fewer than unknowns. Generally speaking, this is an ill-conditioned problem, but we may solve $x$ under some conditions in which $A$ and sparse $x$ are appropriately chosen. The matrix $A$ or a transform of $A$ must obey RIP restriction, which indicates that the random matrix is proper in almost all situations. The signal $x$ must be sparse or sparse under certain basis such as wavelets and discrete cosine transform (DCT):

$$x = \psi x' \tag{2}$$

where $\psi$ is a $n\times n$ transform matrix, and $x'$ is a sparse vector with $n$ dimension. if $x'$ has $k$ nonzero elements and $k \ll n$, we call that the vector $x'$ is k-sparse. The Eq.(1) becomes

$$y = A\psi x' + e \quad \text{or} \quad \begin{cases} y = A'x' + e \\ A' = A\psi \end{cases} \tag{3}$$

We can find an appropriate way to solve Eq.(2) and Eq.(3) by solving the following optimization problem:

$$\min_{x'} \frac{1}{2}\left\| y - A'x' \right\|_2^2 + \tau \left\| x' \right\|_1 \tag{4}$$

where $\tau$ is a parameter weighting the two terms in Eq.(4) which indicate the consistencies of solution with measurement results and prior knowledge of sparse, respectively, and $\|\cdot\cdot\|_p$ stands for $l_p$ norm, defined as $\left(\|x\|_p\right)^p = \sum_{i=1}^{N} |x_i|^p$. For the vector $x'$ with $k$-sparse, it can be accurately reconstructed with $q \geq Ck\log(n/k)$ measurements, in which $n$ is the element number of $x'$ and C is a constant coefficient.[11,23] Howland et al. gave out a simplified formula, $q \geq 4k \approx n/4$.[26] In fact, we usually use $n/3$ as an empirical value in the experiment.

Although in theory there are many measurement matrices that can be selected, in many experiments including ours, only two valued matrices such as "0-1" or "-1-1" can be selected limited by optical components and devices. Most of these matrices obey the Bernoulli distribution. In some specific applications, sparse matrices can also be used, with only a small number of "1" in the matrix.

## 3. Compressive fluorescence spectral imaging

A typical spectral imaging data is a three-dimensional digital array which is organized in the data cube[20]. As shown in Fig.1, the $x$, $y$ represents spatial information and the $\lambda$ represents spectral information. In this data cube, a fixed $\lambda$ determines the spatial image and a fixed $x$-$y$ determines the spectral sequence.

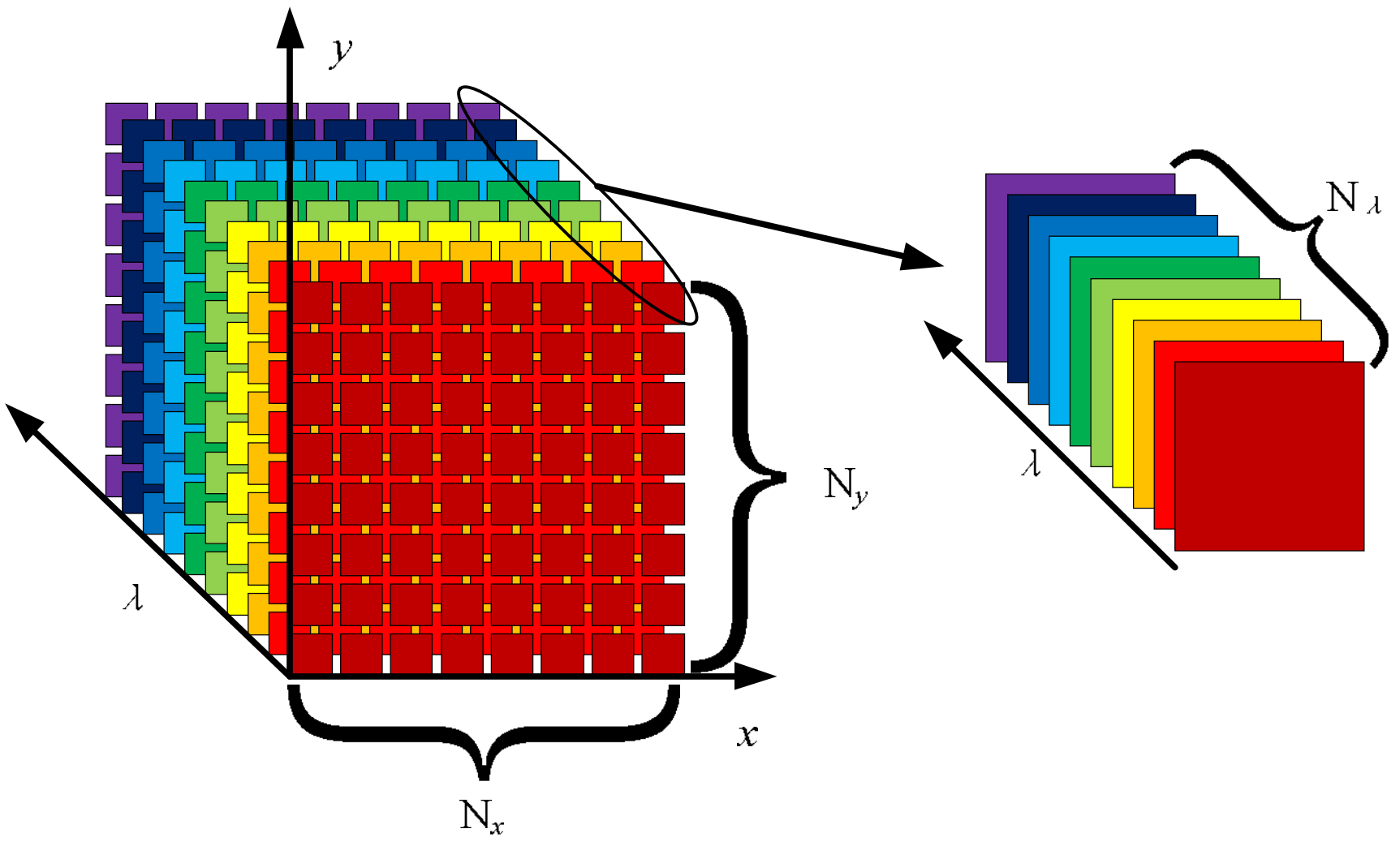

**Fig. 1** Spectral imaging data cube

In traditional methods, we would obtain the data cube through mechanically scanning each spatial point with a spectrometer, which is prone to error and is labor-intensive. Therefore, we developed a compressive fluorescence spectral imaging (CFSI) system for measuring in spatial and spectral domains at the same time without mechanical scan.

With the "0-1" random matrix which obeys the Bernoulli distribution loaded on the digital micromirror device(DMD), usually the random matrix exists about 50% "1" and 50% "0". therefore, our system can collect about 50% light energy from the object to the detector. Compared with mechanical scanning which measures light from just one spatial point per measurement, the CFSI can obtain more energy in each measurement. The CFSI system is shown in Fig.2.

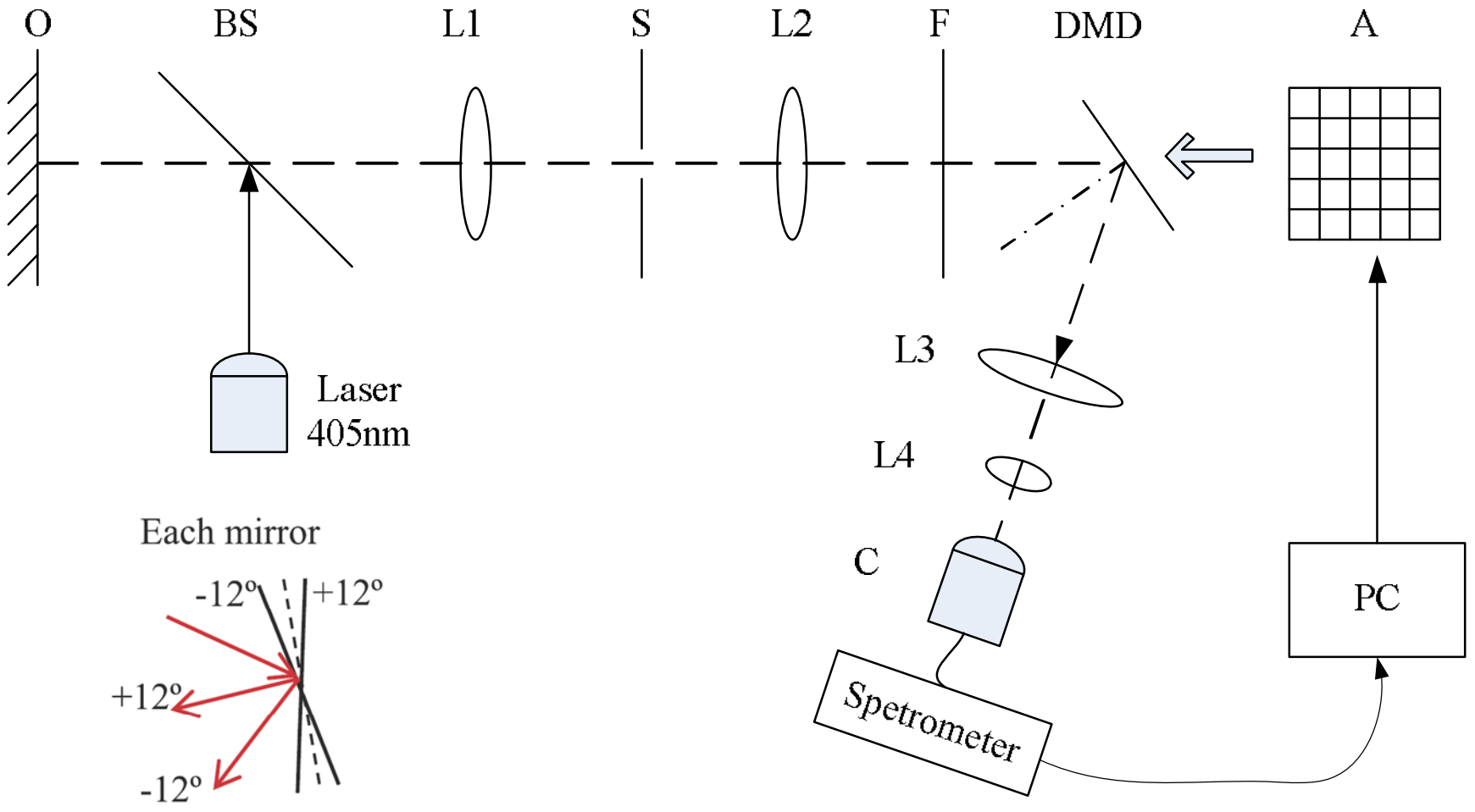

**Fig.2.** CFSI system

With the 405nm laser source exciting an object, the fluorescent image is projected onto the DMD through two lenses, L1 and L2. To simplify the reconstruction, we make a microscopic image of the object on a small part of the DMD. The image has a size of about $1 \times 1mm^2$ and occupies $64 \times 64$ pixels. The whole DMD consists of 1024 × 768 micro-mirrors which independently reflect the light to two directions "0" and "1", according to the "0-1" random matrix loaded on the DMD, where "0" and "1" present -12° and +12° of the micro-mirrors, respectively. "O" is the object prepared with fluorescent materials. "S" is a slit, used to adjust the luminous flux and block stray light. The long wave pass filter "F" blocks the 405nm laser in the imaging light path. Lenses "L3" and "L4" are placed in the "1" direction, collecting light from parts of the image with position coordinates corresponding to "1" in the matrix to the fiber collimator "C". The light is then transferred by the fiber to a spectrometer to measure the spectra. In the experiment, $N$ random matrix $A$ will be loaded on the DMD, making different parts of the object be collected, and $N$ correspondent spectra lines $y$ are given by the spectrometer. The random matrix $A$ and spectra $y$ make up a linear equation as Eq. (5)

$$y(\lambda) = A(x)t(x,\lambda) \tag{5}$$

where $t(x,\lambda)$ is the transmission function of object under different wavelengths $\lambda$, and $x$ is the two-dimensional coordinate of the image on the DMD. Measurement matrix $A$ is independent of $\lambda$ as the reflection of the DMD is uniform for every wavelength within the working range of 400nm-760nm. Therefore, with intensities of each wavelength in the spectra lines and corresponding measurement matrix, the object of different wavelengths can be imaged respectively.

## 4. Experiment

In the experiment we used a commercial spectrometer with spectral range of 200nm to 1100nm and spectral resolution of 1.4nm. First we measured the fluorescence spectra of two types of materials which comprise the object, which are shown in Fig.3a and Fig.3b, respectively. The spectral peak of light purple material is at 468nm and the pink material at 636nm. We made an object with the shape of character "N", as shown in Fig.4. The image on the DMD occupies 64 × 64 pixels which is therefore the resolution of final imaging result. To obtain the spectral images 2468 random matrices were used to drive the DMD, and the spectrometer made a measurement during each state of DMD. Finally, we obtained the data shown in Fig.5. The spectrometer outputs 2468 sets of spectral data with the DMD flipping 2468 times. The spectrums shown in Fig.3a and Fig.3b of light purple and pink Materials were measured by the spectrometer mentioned in the Fig.2. According to the peak values of the spectrums, we reconstructed the image depending on CS using the vertical data at 468nm and 636nm. The CS algorithm we used is Total Variation Augmented Lagrangian Alternating Direction Algorithm(TVAL3) which calculates the sparsest solution in gradient domain under "0-1" random measurement matrix. Fig.6a is the reconstructed image at 468nm and the Fig.6b at 636nm. We add the pixel values of the two images in Fig.6a and Fig.6b together as the fusion image, shown in Fig.6c. As a comparison, we also use the orthogonal matching pursuit (OMP) algorithm to reconstruct the image based on spatial sparse. Fig.7a is the reconstructed image at 468nm and the Fig.7b at 636nm. Fig.7c is the fusion image. Obviously, the image quality of Fig.6 is better than Fig.7 for the different algorithms. Since most of the common images are sparse in the gradient domain, the TVAL3 algorithm is more suitable for image reconstruction.

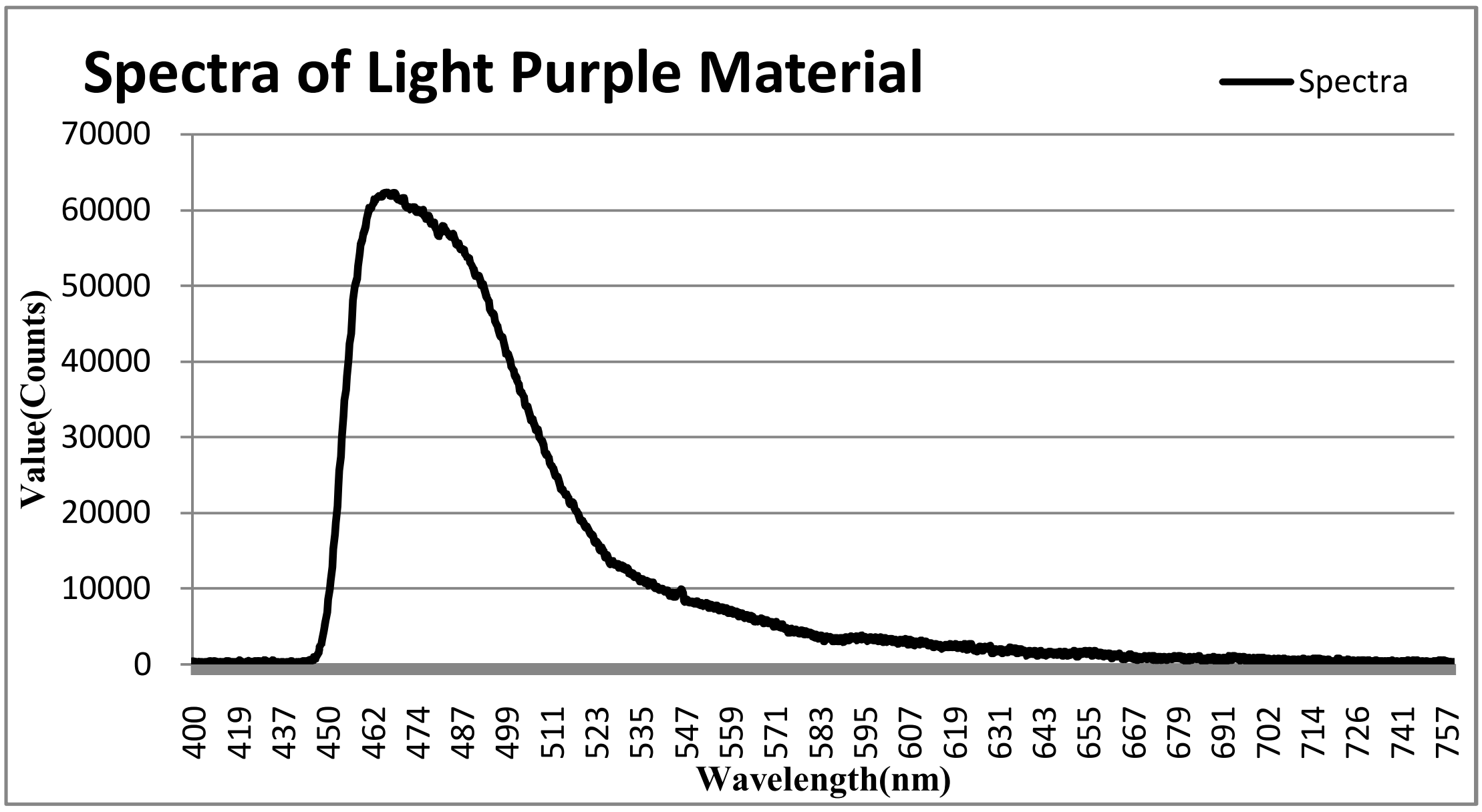

**Fig.3a.** Spectra of Light Purple Material

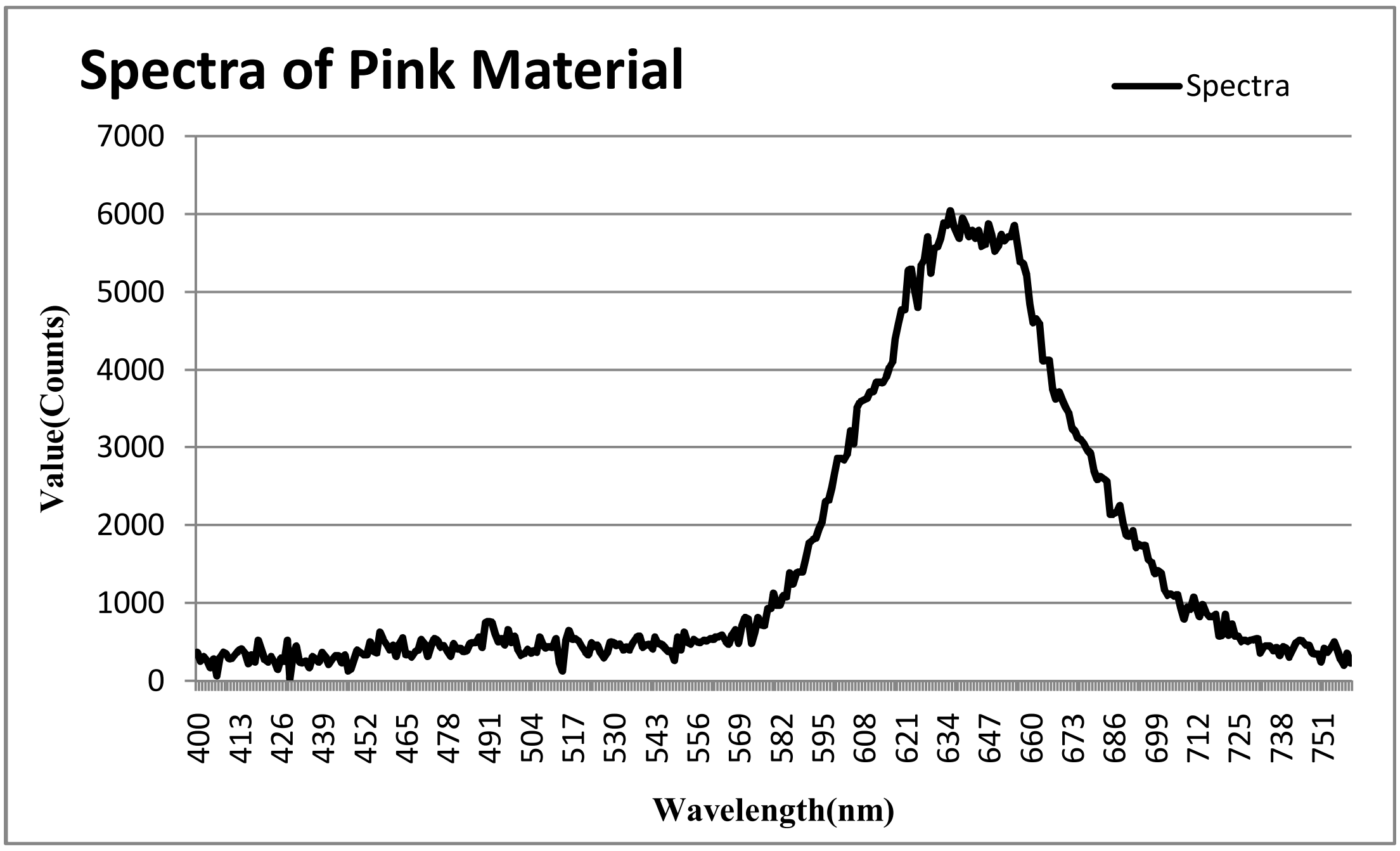

**Fig.3b.** Spectra of Pink Material

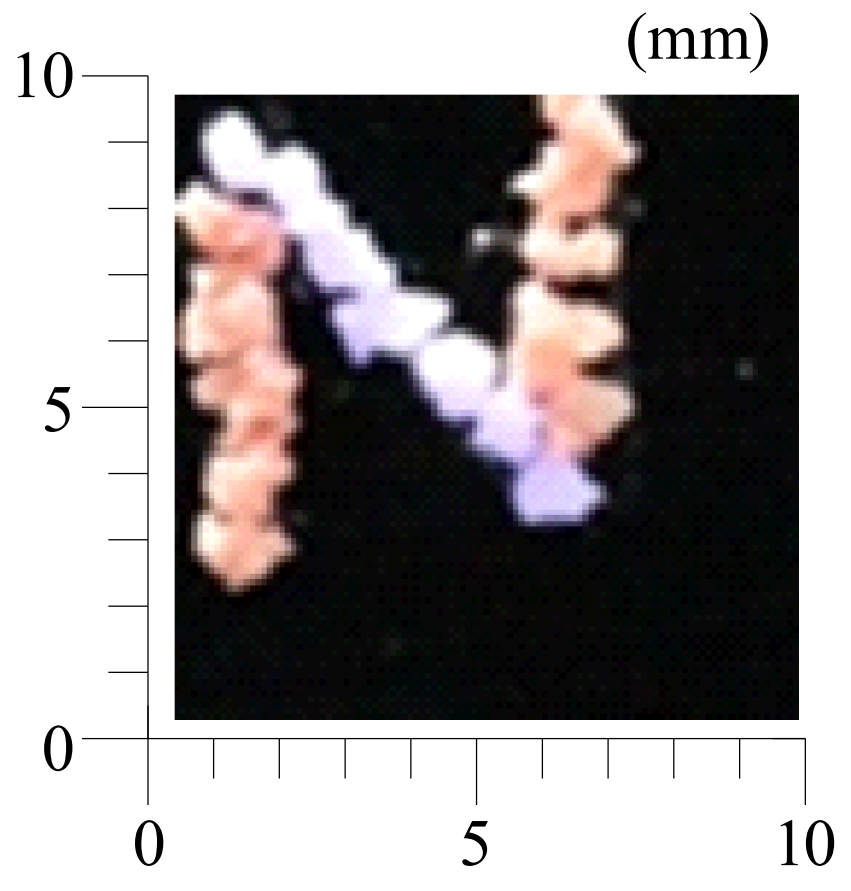

**Fig.4.** The object "N" composed of the fluorescent materials

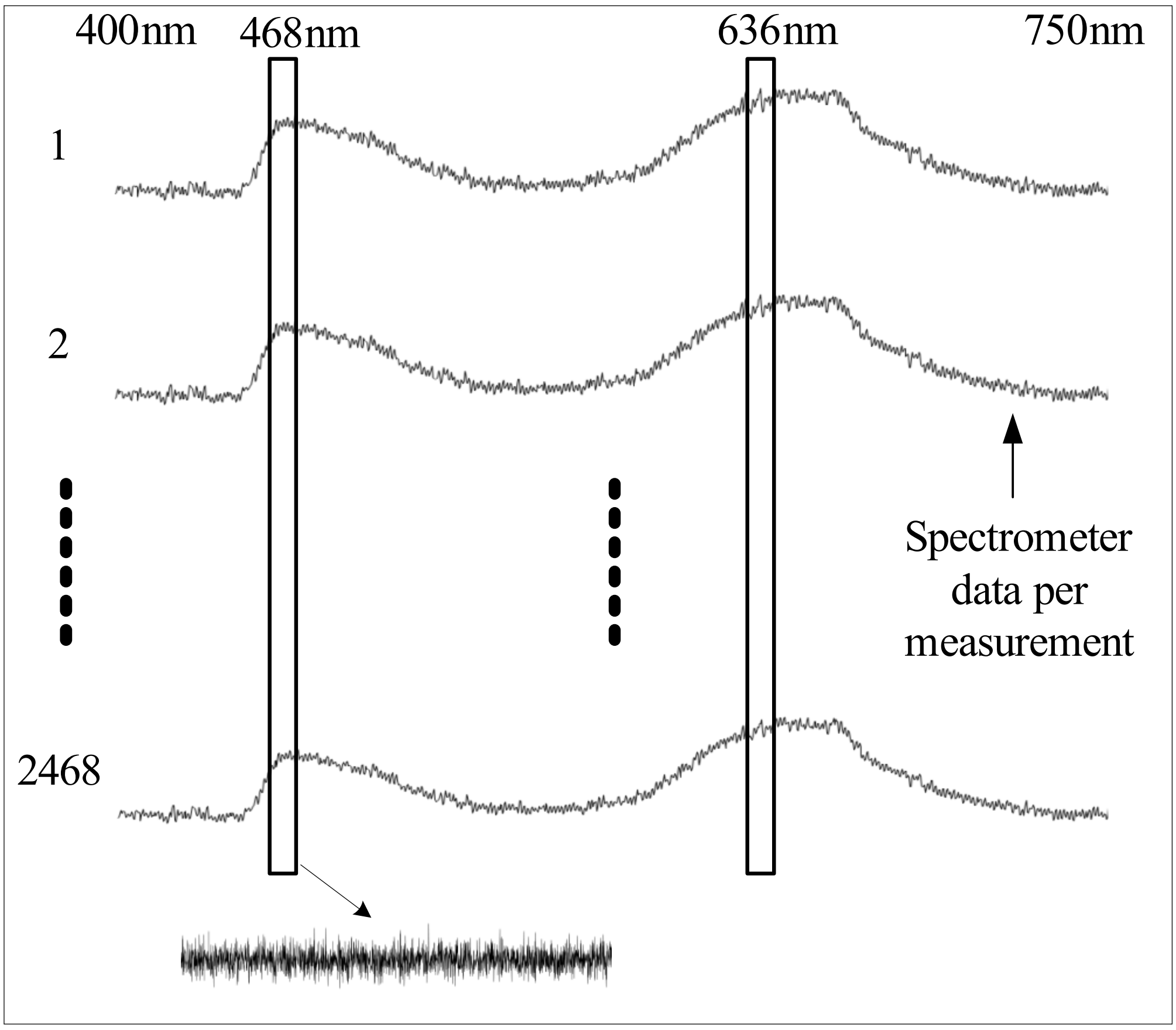

**Fig.5.** The measurement data and application data(only in the visible spectrum)

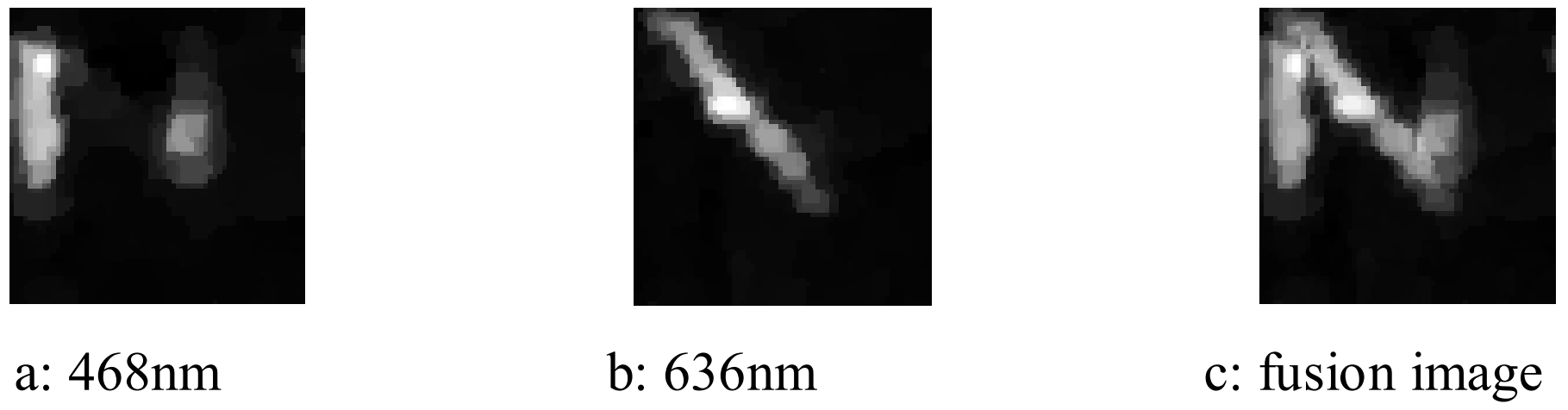

**Fig.6.** Spectral imaging result of the fluorescent materials with TVAL3

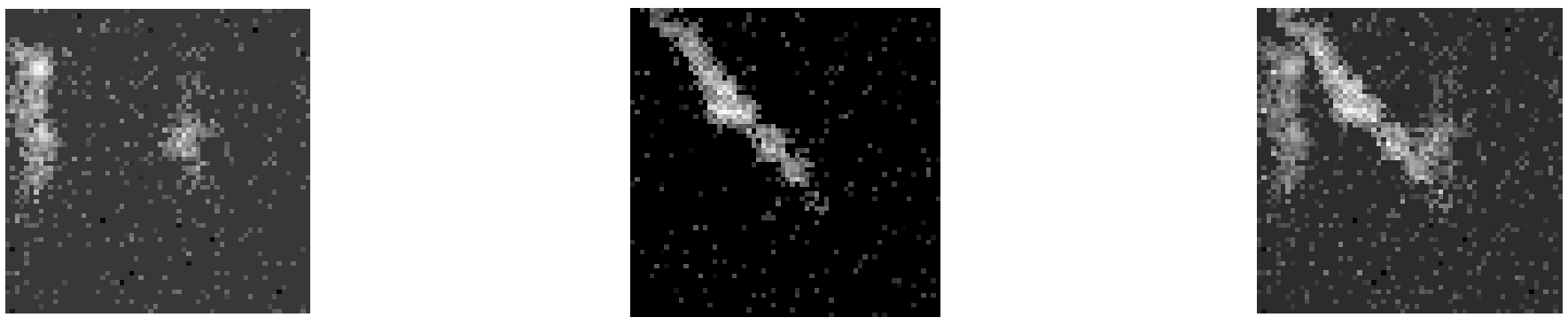

a: 468nm b: 636nm c: fusion image

**Fig.7.** Spectral imaging result of the fluorescent materials with OMP

Indeed, image under every wavelength can be obtained respectively based on the data in our experiment. The spatial and spectral information is measured simultaneously. As the two-dimensional spatial information can be obtained with point measurement based on CS, scanning in neither spatial nor spectral dimensions is required. Besides, the sampling number can be much less than the image pixels, reducing the measurement time compared with traditional spatial scanning, in which the spectrum measurement must be implemented for every single pixel. Therefore, our experiment provides a stable, high effective scheme of spectral imaging.

In the CFSI system, we obtained the image with a size of 64 × 64 pixels with the measurement value of 2468, and the sampling rate was about 60%. Usually, the CS algorithm has a feature that high sampling rate results in image of high quality. As shown in Fig.8, (a) is the CCD image, and (b-g) are the fluorescent material images with 10%, 20%, 30%, 40%, 50% and 60% sampling rates reconstructed by CFSI system using TVAL3.

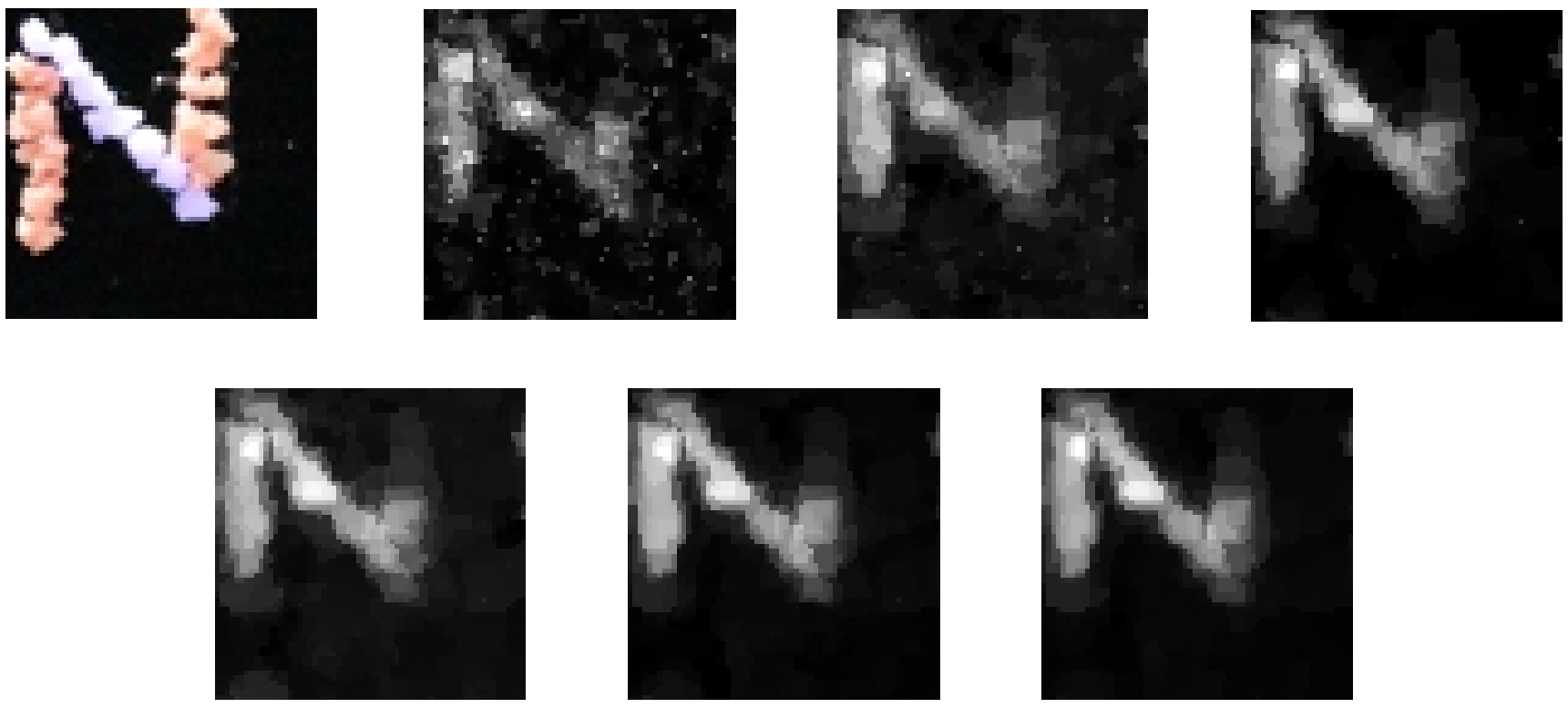

a: CCD image b: 10% c: 20% d: 30% e: 40% f:50% g:60%

**Fig.8.** The image quality with different sampling rates(TVAL3)

Because the image always contains certain structural information and there is correlation between pixels, we evaluated the image using the structural similarity index measurement (*SSIM*) which scales the distortion of structure information by brightness, contrast, and structure. The original image and the reconstructed image were divided into average blocks. The total number of blocks was $B$, whereas $b$ and $b'$ represent the blocks in the original image and the reconstructed image respectively. The equation (6) can be used to calculation the *SSIM*.

$$SSIM(b,b') = l(b,b')^{\alpha} \bullet c(b,b')^{\beta} \bullet s(b,b')^{\gamma} \tag{6}$$

and $l(b,b') = \dfrac{2\mu_b\mu_{b'} + C_1}{\mu_b^2 + \mu_{b'}^2 + C_1}$, $c(b,b') = \dfrac{2\sigma_b\sigma_{b'} + C_2}{\sigma_b^2 + \sigma_{b'}^2 + C_2}$, $s(b,b') = \dfrac{\sigma_{bb'} + C_3}{\sigma_b\sigma_{b'} + C_3}$,

where $\alpha$ , $\beta$ , $\gamma$ are the weights of brightness, contrast and structure, respectively. The $\mu_b$ , $\mu_{b'}$ , $\sigma_b^2$ , $\sigma_{b'}^2$ , $\sigma_b\sigma_{b'}$ are the means, variances and covariances of $b$ and $b'$. The $l(b,b')$ , $c(b,b')$ , $s(b,b')$ are the comparison function of brightness, contrast and structure. Finally, the mean *SSIM*(*MSSIM*) can be calculated by equation (7)[25].

$$MSSIM(x,x') = \frac{1}{B}\sum_{j=1}^{B} SSIM(b_j, b_j^{'}) \tag{7}$$

The result of MSSIM is shown in Fig.9 and $\alpha=\beta=\gamma=1$. If the sampling rate was less than 30%, the quality of reconstructed image was very poor. However, if the rate was more than 30%, the image quality was significantly improved. Indeed, We can easily find that the image quality rises slowly at the sampling rate 30%-60% from Fig.9. We know that higher sampling rate means more sampling time in the experiment. Therefore, in consideration of the practical application based on the compromise strategy, it is recommended to use the sampling rate of just over 30%.

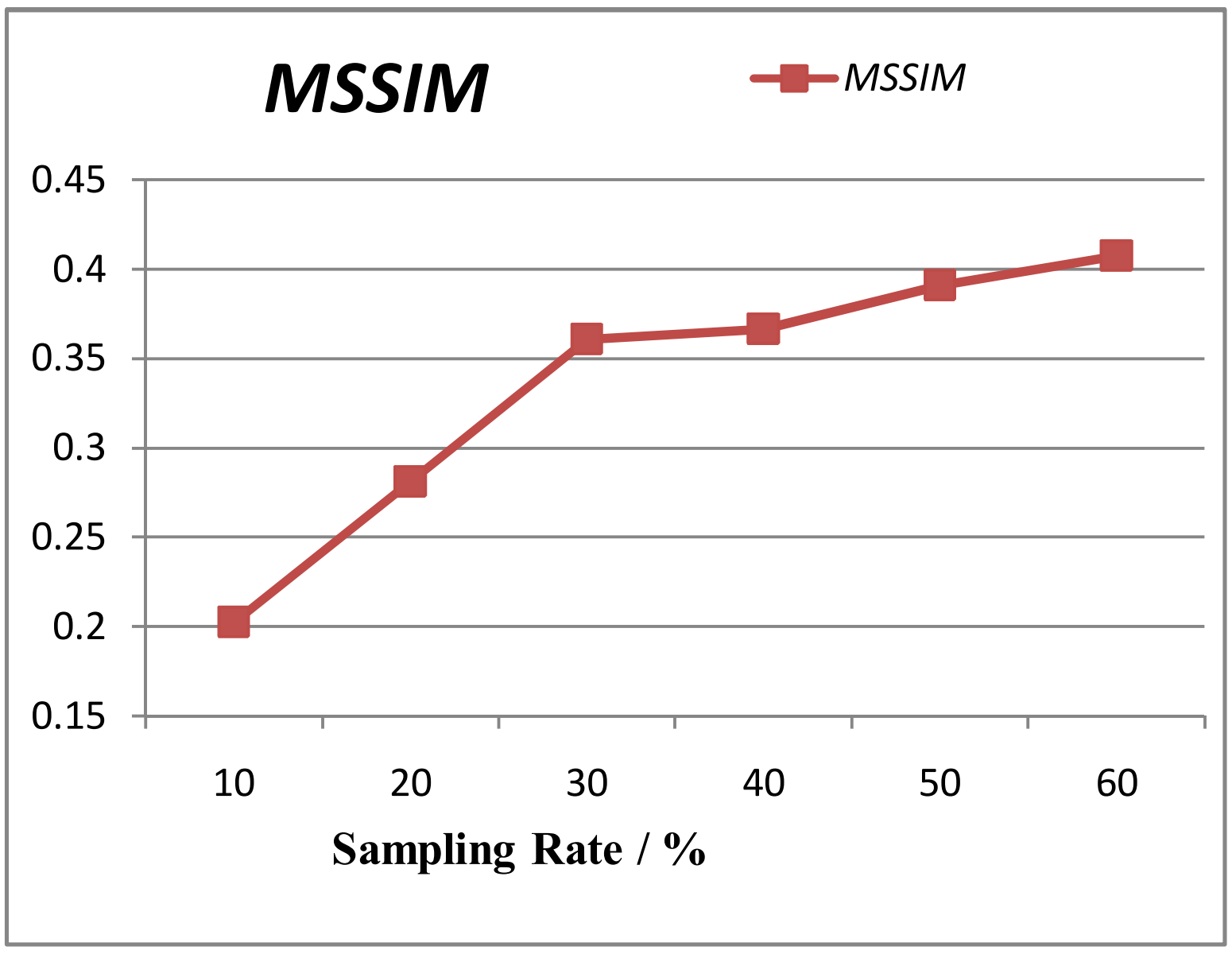

**Fig.9.** Image quality assessment with *MSSIM*

## 5. Conclusions

In summary, we have demonstrated the spectral and spatial imaging method with a spectrometer without any mechanical scanning in the CFSI system. With the DMD modulation providing the spatial resolution, we can obtain spectral images within the wavelength range of the spectrometer simultaneously. The spectral resolution depends on the optical system and the spectrometer. Although the quality of image is not perfect, it is possible to be improved with the development of CS algorithm. Our method provides a stable, high speed scheme of fluorescence spectral imaging, which is anticipated to be useful in biomedicine, astronomy, material science, and other related fields.